\documentclass[a4paper,11pt]{article}
\usepackage{pos}
\usepackage{subcaption}

\title{A Decade of Black-Hole X-ray Binary Transients}
\ShortTitle{BH XRB Transients}

\author*[a,b]{P.A. Charles}       
                       
\author[b,c,d]{D.A.H. Buckley}

\author[c]{E. Kotze}

\author[c]{M. Kotze}

\author[c]{J.K. Thomas}

\author[a]{P. Gandhi}

\author[a]{J.A. Paice}

\author[e,f]{J.-P. Lasota}

\author[g]{J.H. Matthews}

\author[h]{J.F. Steiner}

\affiliation[a]{Department of Physics and Astronomy, University of Southampton, Southampton SO17 1BJ, UK}

\affiliation[b]{Department of Physics, University of the Free State, PO Box 339, Bloemfontein 9300, South Africa}

\affiliation[c]{South African Astronomical Observatory, Observatory Road, Observatory, 7925, Cape Town, South Africa}

\affiliation[d]{Department of Astronomy, University of Cape Town, Private Bag X3, Rondebosch 7701, South Africa}

\affiliation[e]{Institut d’Astrophysique de Paris, CNRS et Sorbonne Université, UMR 7095, 98bis Bd Arago, F-75014 Paris, France}

\affiliation[f]{Nicolaus Copernicus Astronomical Center, Polish Academy of Sciences, Bartycka 18, 00-716 Warsaw, Poland}

\affiliation[g]{Institute of Astronomy, University of Cambridge, Madingley Road, Cambridge, CB3 0HA, UK}

\affiliation[h]{Center for Astrophysics, Harvard-Smithsonian, 60 Garden St, Cambridge, MA 02138, USA}

\emailAdd{P.A.Charles@soton.ac.uk}

\abstract{The last decade has seen a significant gain in both space and ground-based monitoring capabilities, producing vastly better coverage of BH X-ray binaries during their (rare) transient events. This interval included two of the three brightest X-ray outbursts ever observed, namely V404\,Cyg in 2015, and MAXI\,J1820+070 in 2018, as well as the outburst of Swift\,J1357.2-0933, the first such system to show variable period optical dipping. There are now superb multi-wavelength archives of these outbursts, both photometric and spectroscopic, that show substantial outflows in the form of jets and disc winds, and X-ray spectroscopy/timing that reveals how the inner accretion disc evolves. The ground-based AAVSO optical monitoring of the MAXI\,J1820+070 event was the most extensive ever obtained, revealing periodic variations that evolved as it approached its state transition.  These modulations were of an amplitude never seen before, and suggested the development of an irradiation-driven disc warp that persisted through the transition.  All these results have demonstrated the power of extensive multi-wavelength photometric and spectroscopic monitoring on all time-scales.}

\FullConference{
  *** High Energy Astrophysics in Southern Africa (HEASA2021) ***\\
  *** 13 - 17 September, 2021 ***\\
  *** Online ***
}

\begin{document}
\maketitle

\section{Introduction}

\begin{figure}[htb]
\centering
\setkeys{Gin}{width=\linewidth}
\begin{subfigure}{0.5\textwidth}
\scalebox{1.0}{\includegraphics{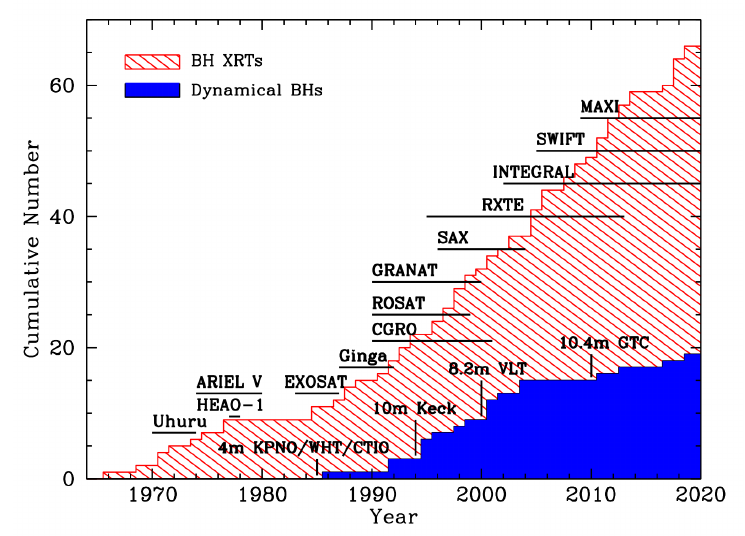}}
\caption{Discovery of BHXTs since the start of X-ray astronomy (updated version of \citep{BlackCat16})}
\label{fig:BHXRTs-Timeline}
\end{subfigure}
\hfil
\begin{subfigure}{0.45\textwidth}
\scalebox{1.0}{\includegraphics{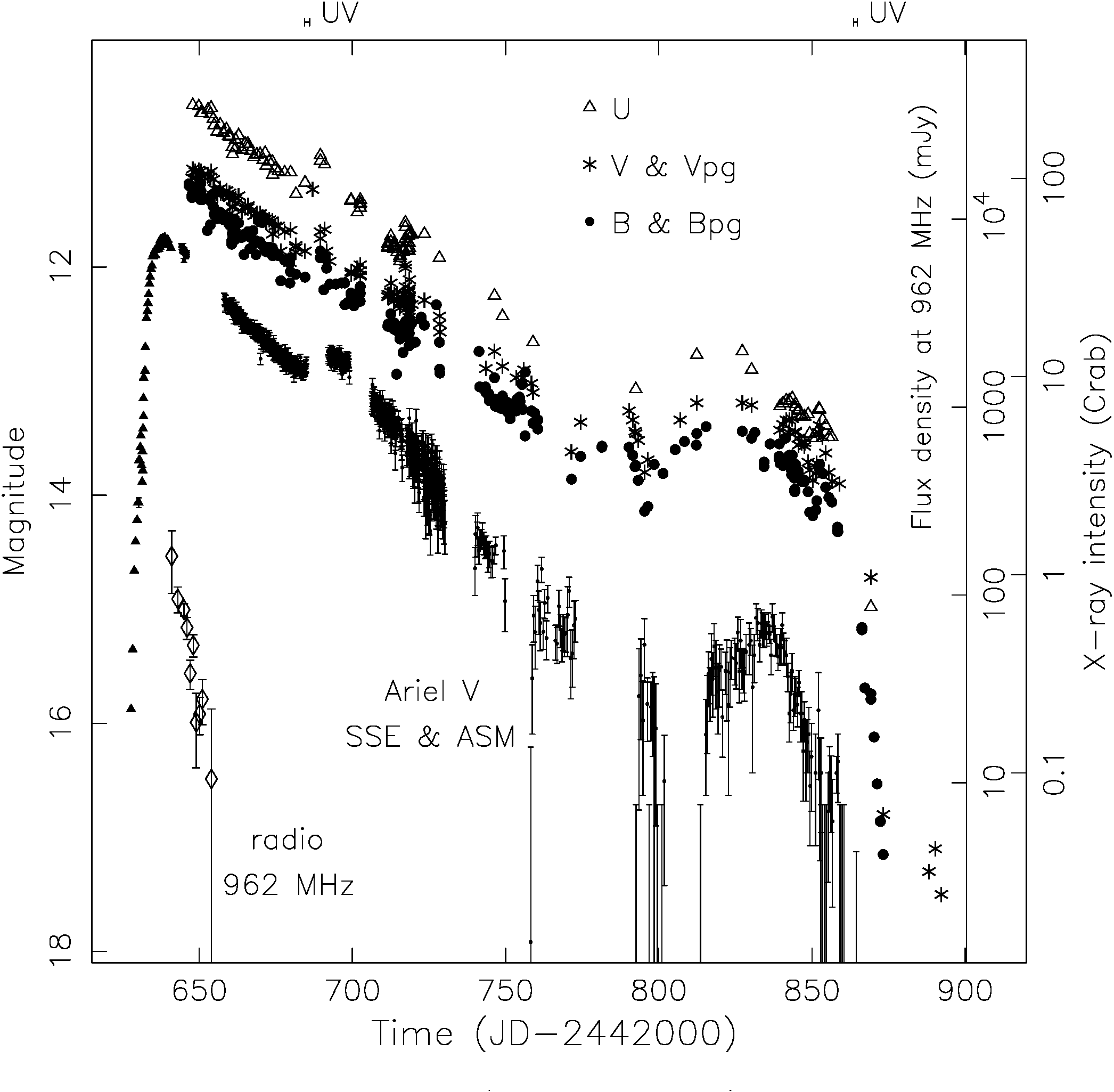}}
\caption{Optical/X-ray light-curves of the 1975 outburst of A0620-00 \citep{Kuulkers98}}.
\label{fig:A0620LCs}
\end{subfigure}
\caption{Timeline of BHXTs discovery and 1975 outburst of A0620-00}
\end{figure}

While the first Black-Hole X-ray Transient (BHXT) was discovered via the early rocket sky-surveys of the 1960s, their numbers grew slowly, only increasing at the current $\sim$2/yr since the 1990s and the advent of more sensitive, all-sky X-ray monitoring satellite capability (Figure \ref{fig:BHXRTs-Timeline}).  BHXTs are all highly evolved, low-mass X-ray binaries (LMXBs) containing a low-mass (usually $<$1\,M$_\odot$) donor star transferring material via Roche-lobe overflow into an accretion disc that surrounds the compact object, of which A\,0620-00 (Nova Mon 1975) is the class prototype (Figure \ref{fig:A0620LCs}, \citep{Kuulkers98}).  Their orbital periods are $\sim$hours--days, but BHXTs are only discovered through their rare, highly luminous X-ray outbursts that occur when the disc accumulates sufficient material that it enters a hot, viscous state, enabling high levels of mass to be accreted onto the compact object \citep{Lasota01}.  This can be a neutron star (usually indicated by the presence of type 1 X-ray bursts, or X-ray pulsations) or a BH, the latter requiring additional evidence in the form of a spectroscopic orbit for the donor, in order to provide a dynamical measurement of the compact object mass \citep{Casares17}.  Through this route, BHXTs currently provide the {\it only} source of accurate, stellar-scale BH masses, but it is difficult, as it can only be done in quiescence, and the low-mass donors are faint (which is why, in Figure \ref{fig:BHXRTs-Timeline}, increases in X-ray sensitivity are detecting fainter BHXTs, but confirming many of them dynamically will require future, extremely-large optical telescopes).  Nevertheless, by studying their remarkable range of variability (on all timescales) which is seen at all wavelengths when approaching, during and after outburst, we can gain insights into the physics of high mass-transfer rate accretion and ejection processes, and that is the main subject of this review.

Here we address the last decade, where the continuous, all-sky X-ray monitoring provided by the Swift and MAXI missions has revealed typically a half-dozen BHXTs in outburst during each year.  However, from this database we focus on three BHXTs that stand out: V404\,Cyg and MAXI\,J1820+070 are (together with A\,0620-00) the brightest X-ray sources ever detected (their key properties are summarised in Table \ref{tab:BHXT-props-table}, with precise distances now provided by {\it Gaia} \citep{Gandhi19}); and the third is Swift\,J1357.2-0933 (hereafter ``J1357''), which exhibited absolutely unique outburst behaviour.

{\small
\begin{table*}
	\centering
	\caption{Key Properties of the 3 Brightest BHXTs $^1$}
	\label{tab:BHXT-props-table}
\begin{tabular}{l c c c c c}
\hline
\textbf{Property} & \textbf{A\,0620-00} & \textbf{V404\,Cyg} & \textbf{MAXI\,J1820+070}  \\
\hline
$d$ (kpc) & 1.6 & 2.4 & 3.0 \\
$P_{\rm orb}$ (d) & 0.32 & 6.47 & 0.68 \\
$f(M)$ (M$_\odot$) & 2.8 & 6.1 & 5.2 \\
$q (= M_{\rm 2}/M_{\rm X})$ & 0.074 & 0.067 & 0.072 \\
donor & K2--7\,V & K3\,III/IV & K4/5\,IV/V \\
$i$ ($^\circ$) & 51 & 56--67 & 63--81 \\
$a$ (R$_\odot$) & 3.7 & 31.5 & 6.8 \\
$M_{\rm X}$ (M$_\odot$) & 6.6 & 9--12 & 6--9 \\
$L\rm{_X^{max}}$ (2--10\,keV) (10$^{38}$\,erg\,s$^{-1}$) & 7.8 & 1.7 & 1.6 \\
\hline
\end{tabular}
\begin{flushleft}
$^{1}$ BlackCat \citep{BlackCat16}, \citep{Torres20}
\end{flushleft}
\end{table*}
}

\section{2011 -- Swift~J1357.2-0933: Period-changing optical dips}

Discovered with the Swift BAT in 2011, this high latitude (b=+50$^\circ$) XRT was initially thought to be one of the very faint ($L_X \sim$10$^{35}$ erg\,s$^{-1}$) transients, on the basis of its optical brightness in quiescence being associated with the donor star \citep{ArmasPadilla14}.  However, quiescent spectroscopy \citep{Torres15} subsequently revealed only a disc spectrum, with no sign of the donor, indicating that it must therefore be at a much greater distance ($d>$6\,kpc) and hence intrinsically much more luminous.

\begin{figure}[htb]
\centering
\setkeys{Gin}{width=\linewidth}
\begin{subfigure}{0.45\textwidth}
\scalebox{1.0}{\includegraphics{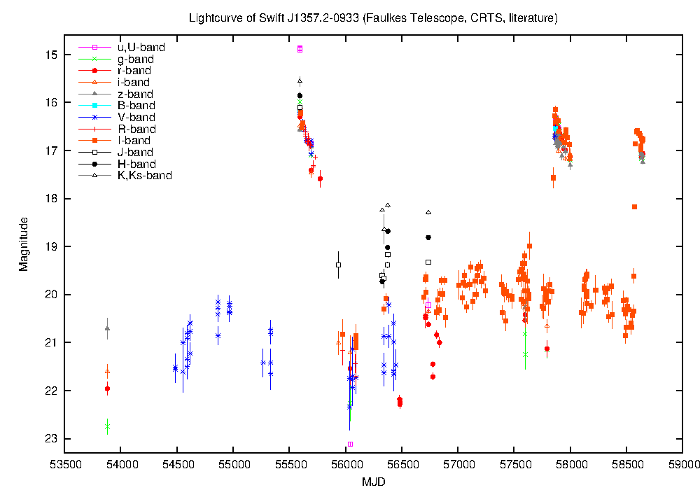}}
\caption{Long-term UV/opt/IR light-curve of J1357 from the XB-News monitoring program (\citep{Russell19}, \citep{Goodwin20}), showing its discovery in 2011 (MJD$\sim$55500) and subsequent outbursts in 2017 and 2019.}
\label{fig:J1357optlc}
\end{subfigure}
\hfil
\begin{subfigure}{0.45\textwidth}
\scalebox{1.0}{\includegraphics{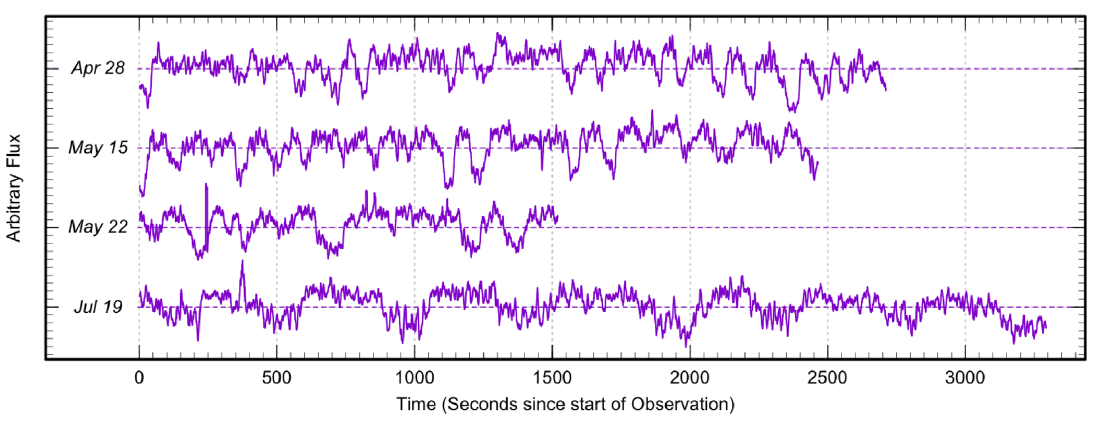}}
\caption{SALT/RSS fast photometry of J1357 during the 2017 outburst \citep{Paice19}.}
\label{fig:J1357lcs}
\end{subfigure}
\caption{Optical light-curves of Swift\,J1357.2-0933}
\end{figure}

\begin{figure}[htb]
\centering
\setkeys{Gin}{width=\linewidth}
\begin{subfigure}{0.45\textwidth}
\scalebox{0.9}{\includegraphics{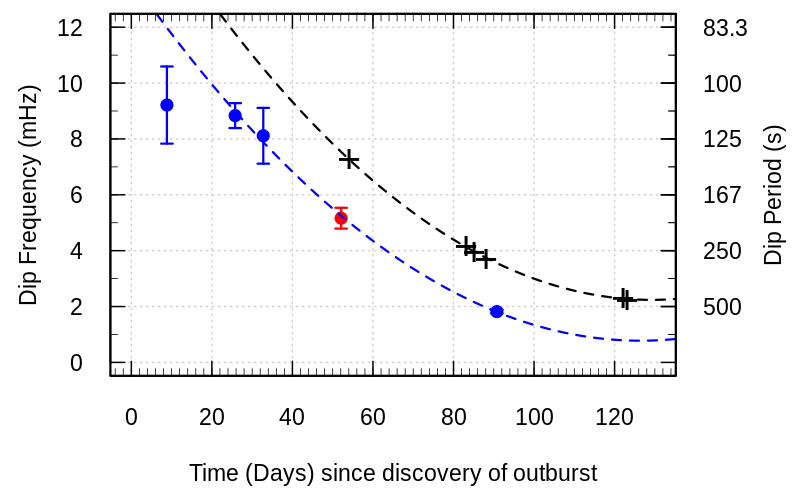}}
\caption{Optical dip frequency evolution during J1357's 2011 (black line) and 2017 (blue line) outbursts \citep{Paice19}.}
\label{fig:J1357dipfreq}
\end{subfigure}
\hfil
\begin{subfigure}{0.45\textwidth}
\scalebox{1.0}{\includegraphics{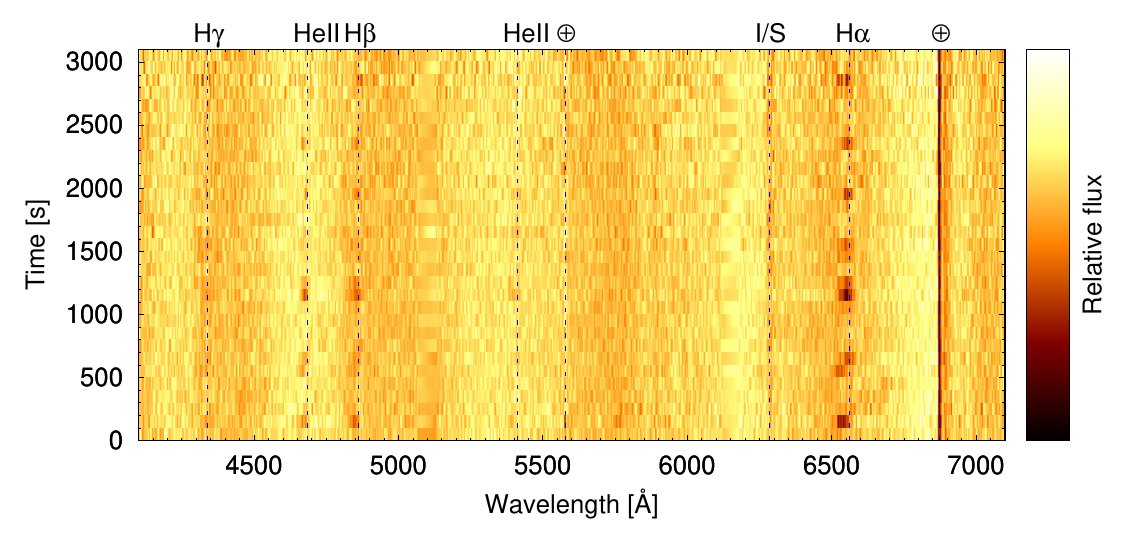}}
\caption{SALT/RSS time-resolved spectroscopy of J1357 in 2017 reveals the Balmer and HeII absorption features associated with the $\sim$500s dips \citep{Charles19}.}
\label{fig:J1357-2dspec}
\end{subfigure}
\caption{Dip-frequency evolution and spectra of Swift\,J1357.2-0933}
\end{figure}

Its most unusual property during the 2011 outburst was that its relatively bright optical counterpart (Figure \ref{fig:J1357optlc}) exhibited periodic dips in the light-curve, but the period of these dips evolved during the course of the outburst from $\sim$2--8 mins \citep{CorralSantana13}!  They attributed this property to an inner disc torus that moved slowly away from the compact object as the outburst proceeded.  Their H$\alpha$ spectroscopy indicated a 2.8\,hr orbital period, making J1357 one of the shortest $P_{\rm orb}$ BHXTs known.  The optical dipping was found to be present again in the 2017 outburst (Figure \ref{fig:J1357lcs}, \citep{Paice19}), and the period evolution in both outbursts (see Figure \ref{fig:J1357dipfreq}), shows very similar trends.

\begin{figure}[htb]
\centering
\setkeys{Gin}{width=\linewidth}
\begin{subfigure}{0.45\textwidth}
\scalebox{1.0}{\includegraphics{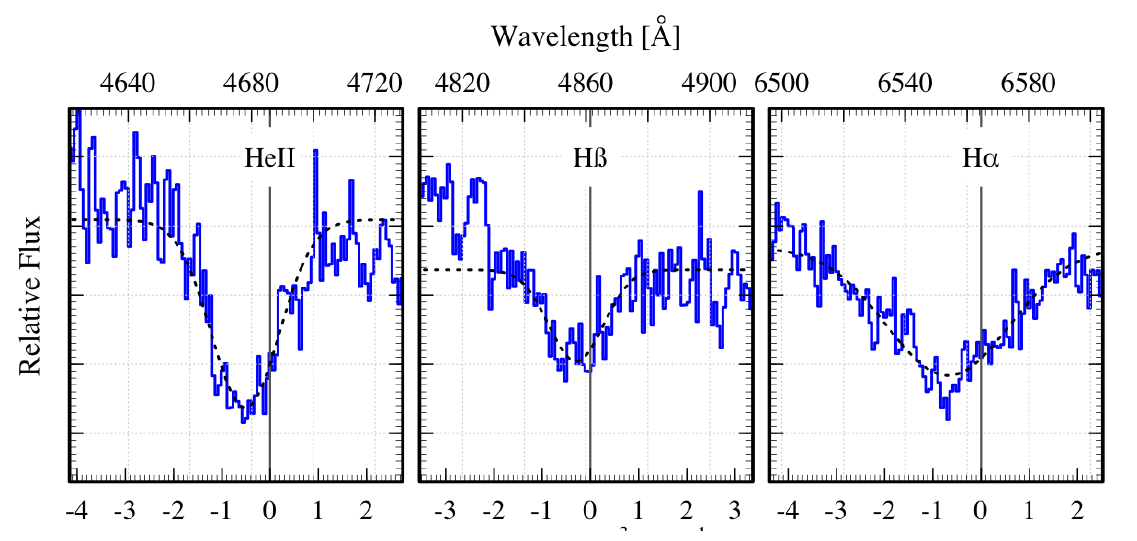}}
\caption{Mean dip optical line profiles for HeII, H$\beta$ and H$\alpha$ in J1357, showing that they are all blue-shifted by $\sim$600 km\,s$^{-1}$ \citep{Charles19}.}
\label{fig:J1357profiles}
\end{subfigure}
\hfil
\begin{subfigure}{0.45\textwidth}
\scalebox{0.8}{\includegraphics{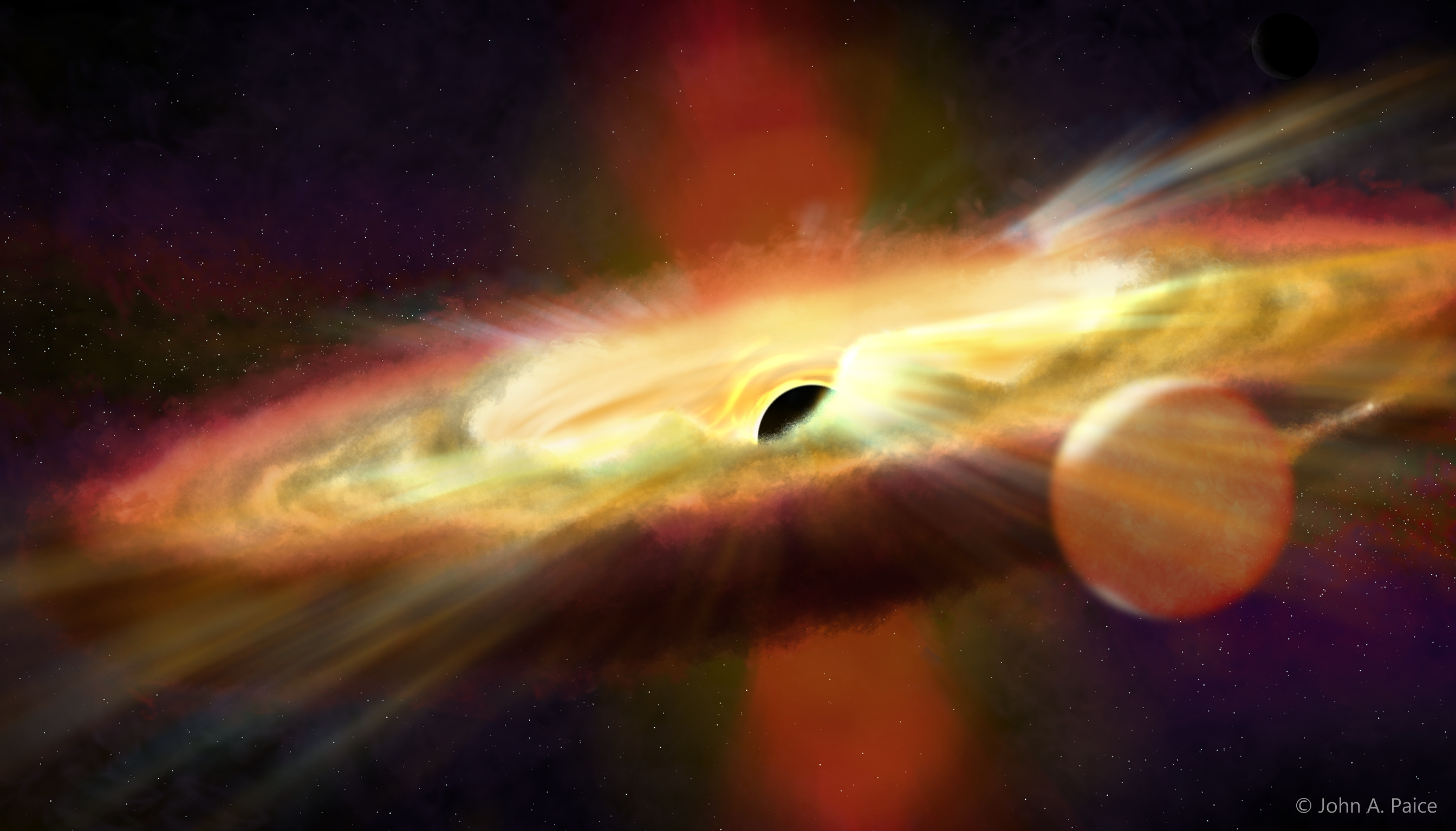}}
\caption{Schematic of J1357 showing an outflowing disc wind from an inner ring (image by John Paice).}
\label{fig:J1357schematic}
\end{subfigure}
\caption{Mean dip line profiles and schematic of Swift\,J1357.2-0933}
\end{figure}

The large aperture of SALT, combined with its fast detectors, made it possible to perform time-resolved spectroscopy through the dips (Figure \ref{fig:J1357-2dspec}).  This revealed the presence of Balmer and HeII absorption features that vary with the dip period \citep{Charles19}, see also \citep{JimenezIbarra19}.  What is most remarkable here is that HeII in absorption had {\it never} been seen in any LMXB before!  Furthermore, the mean dip line profiles (Figure \ref{fig:J1357profiles}) were clearly all blue-shifted by $\sim$600\,km\,s$^{-1}$ relative to the systemic velocity. This led to a model of an outflowing disc wind, which required densities $\sim$10$^{13-14}$\,cm$^{-3}$ and $L_{\rm X}>$10$^{36}$\,erg\,s$^{-1}$ in order to be able to produce HeII in {\it absorption} \citep{Charles19}.  This also requires a high inclination viewing angle, as shown schematically in Figure \ref{fig:J1357schematic}.  The absence of X-ray eclipses, in spite of the inclination, is due to the very compact nature of the donor given the short $P_{\rm orb}$.  But this still requires a very specific viewing angle for J1357, hence accounting for its unique observational properties.

\section{2015 -- V404\,Cyg: Winds and outflows}

\begin{figure}[htb]
\centering
\setkeys{Gin}{width=\linewidth}
\begin{subfigure}{0.45\textwidth}
\scalebox{0.8}{\includegraphics{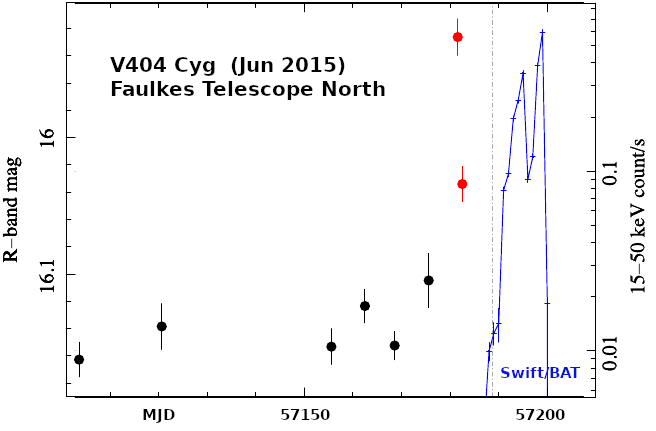}}
\caption{Faulkes Telescope R-band monitoring of V404\,Cyg in the lead-up to the June 2015 X-ray eruption that is marked with the dashed line (Swift/BAT hard X-ray light-curve shown in blue, adapted from \citep{Bernardini16}).}
\label{fig:v404preoutburst}
\end{subfigure}
\hfil
\begin{subfigure}{0.45\textwidth}
\scalebox{1.0}{\includegraphics{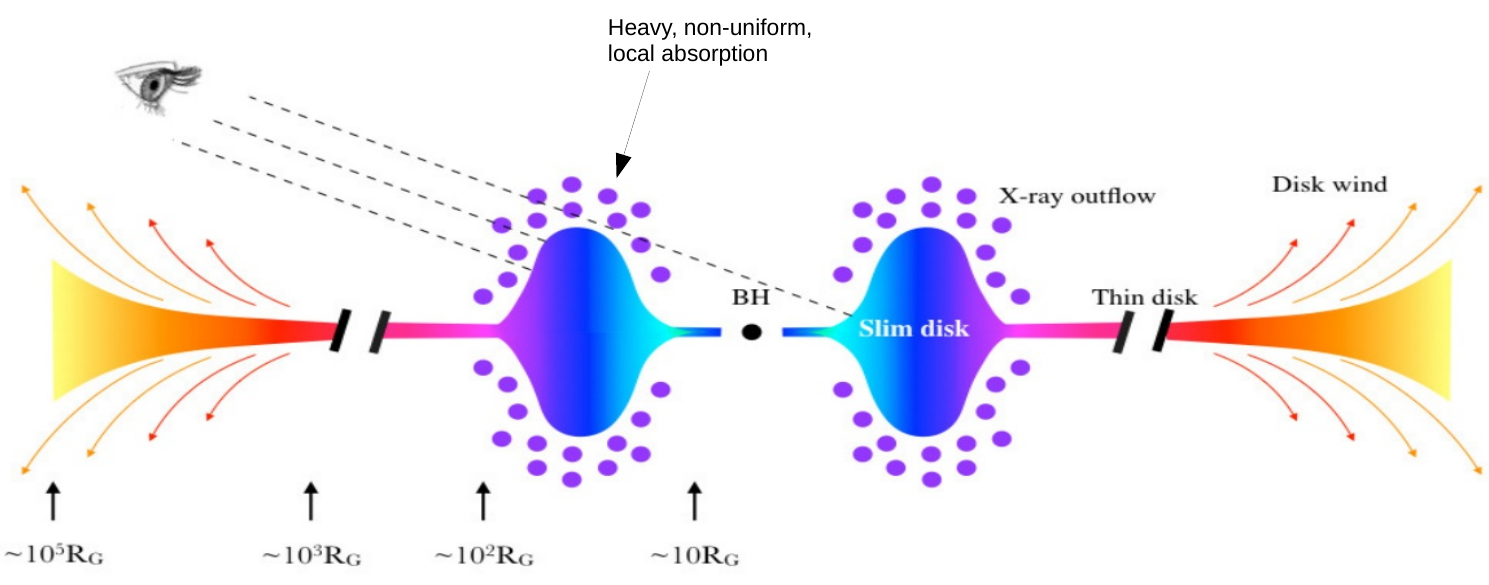}}
\caption{V404\,Cyg schematic showing regions of variable, high absorption from an inner torus within a few hundred gravitational radii ($R_{\rm G}$) of the BH (from \citep{Motta17}).}
\label{fig:v404-Sara-schematic}
\end{subfigure}
\caption{The 2015 outburst of V404 Cyg}
\end{figure}

Discovered through its huge X-ray outburst in 1989 as GS2023+338, its optical identification (V404\,Cyg) was already known as Nova Cyg 1938, thereby demonstrating that it had recurrent outbursts on timescales of $\sim$25--50\,yrs.  So, after returning to quiescence in 1990, V404\,Cyg has been monitored extensively ever since, and this led to the first optical detection of activity about a week prior to its 2015 X-ray outburst (the two points plotted in red in Figure \ref{fig:v404preoutburst}, \citep{Bernardini16}).  This small ($\sim$0.1\,mag) increase in brightness was interpreted through application of the DIM (Disc Instability Model) to BHXTs \citep{Lasota01} as the inner accretion disc radius is beginning to decrease, accompanied by an increase in H$\alpha$ emission and velocity width, which was observed just 13 hours before the X-ray turn-on.

The subsequent X-ray eruption was even brighter than in 1989, but accompanied by the same enormous ($\geq\times$10$^3$) rapid ($\sim$mins) intensity fluctuations.  The relatively short outburst in June 2015 (barely 2 weeks) was extensively monitored by Swift (and other satellites), which allowed for detailed coverage of the X-ray spectral variations associated with these huge intensity fluctuations.  This revealed very large variations in (local) X-ray column density (from a few $\times$10$^{21}$cm$^{-2}$ to $>$10$^{24}$cm$^{-2}$, and led \citep{Motta17} to propose the scenario shown in Figure \ref{fig:v404-Sara-schematic}, citing a possible super-Eddington rate of accretion as the cause of this behaviour.  

\begin{figure}[htb]
\centering
\setkeys{Gin}{width=\linewidth}
\begin{subfigure}{0.45\textwidth}
\scalebox{1.0}{\includegraphics{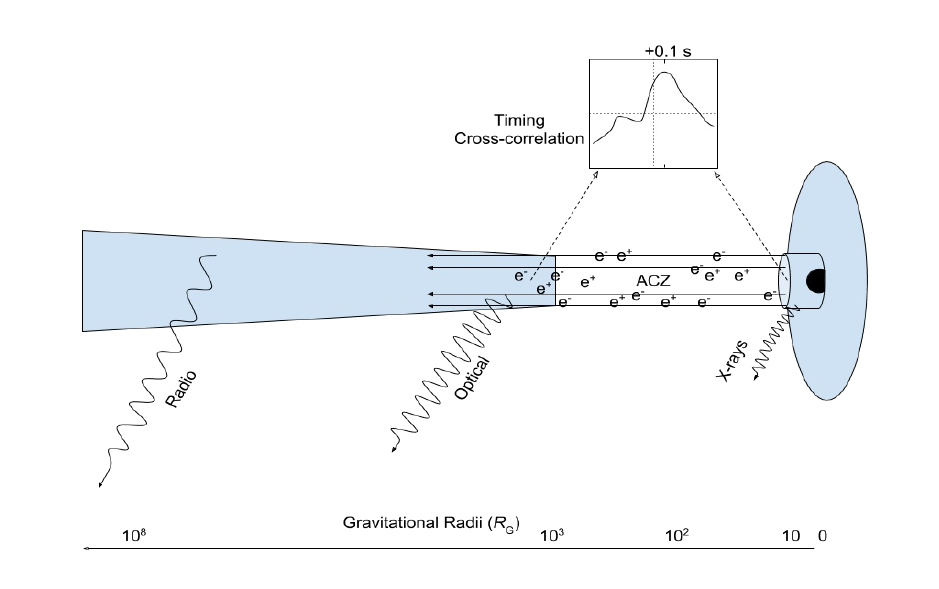}}
\caption{Schematic of inner jet acceleration zone showing X-ray and optical emitting regions separated by $\sim$10$^3R_{\rm G}$, accounting for the 0.1s optical-X-ray delay observed (box) \citep{Gandhi17}.}
\label{fig:v404-jet-schematic}
\end{subfigure}
\hfil
\begin{subfigure}{0.45\textwidth}
\scalebox{0.9}{\includegraphics{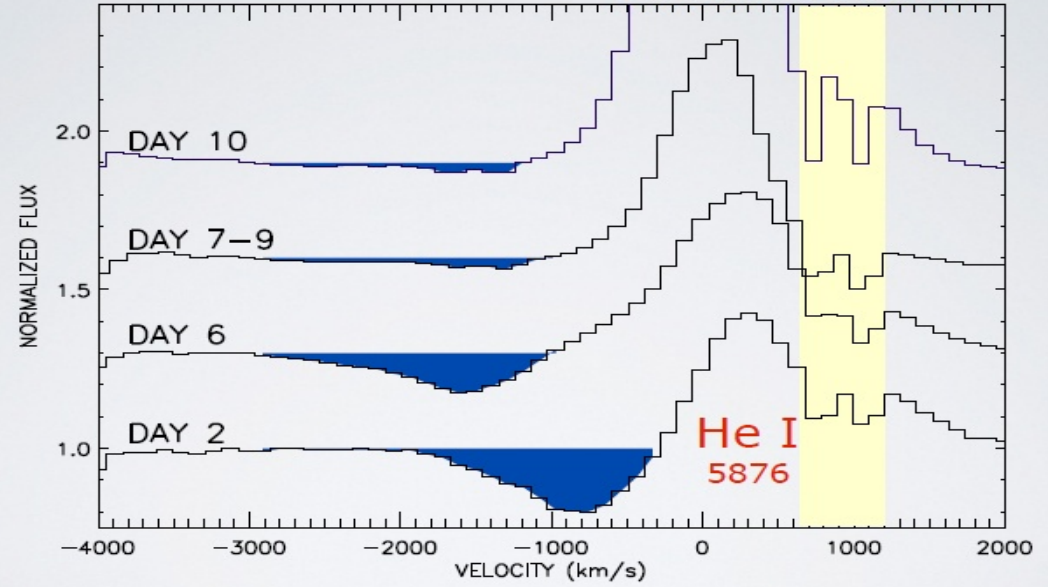}}
\caption{HeI\,$\lambda$5876 P\,Cyg profiles observed during X-ray flaring intervals in V404\,Cyg's 2015 outburst \citep{MunozDarias16}.}
\label{fig:v404-HeI}
\end{subfigure}
\caption{Jet and winds from V404\,Cyg}
\end{figure}

X-ray/optical correlated variability was expected to be taking place with time lags between them of $\sim$35--75\,s (within the disc and heated face of the donor, since the binary separation is 31.5R$_\odot$), or much faster ($\sim$0.1--1\,s) variations associated with the launch of X-ray/radio jets and inner disc winds.  In high-speed, multi-band ULTRACAM photometry, simultaneous with NuSTAR,  intense, rapid ($<$24\,ms) and very red flaring activity was found that lagged the X-rays by $\sim$0.1\,s \citep{Gandhi17}, as shown in Figure \ref{fig:v404-jet-schematic}.  The red colour meant that this could not be thermal reprocessing of X-rays, and so it was interpreted as synchrotron emission in a relativistic jet, with the optical emanating from a height of $\sim$10$^3 R_{\rm G}$, thereby constraining the size of the inner jet acceleration zone.  Similar variations have been seen from GX339-4 and MAXI\,J1820+070.

Time-resolved optical spectroscopy was also undertaken through X-ray/radio flaring episodes \citep{MunozDarias16}, producing the P\,Cyg profiles in HeI\,$\lambda$5876 (Figure \ref{fig:v404-HeI}).  Their strongest feature (on Day 2) was directly associated with an X-ray flare, and was explained as the launching of a strong disc wind, which was sustained for much of the outburst.  But this had to be at sufficiently large disc radii for the He not to be ionised.  The outflow velocity was $\sim$0.01$c$ and carried away a mass $>$10$^{-8}$M$_\odot$, and was thereby likely responsible for curtailing the outburst \citep{MunozDarias16}.  Interestingly, \citep{AlfonsoGarzon18} have found X-ray/optical flaring behaviour during the outburst with lags greater than the light-travel time across the system (i.e. $>$2 mins), which provides strong evidence for the existence of substantial ejected material from V404\,Cyg via winds, as well as that seen via radio jets \citep{Tetarenko17}.

\section{2018 -- MAXI\,J1820+070: Warped, precessing disc}

\begin{figure}[htb]
\centering
\setkeys{Gin}{width=\linewidth}
\begin{subfigure}{0.45\textwidth}
\scalebox{1.0}{\includegraphics{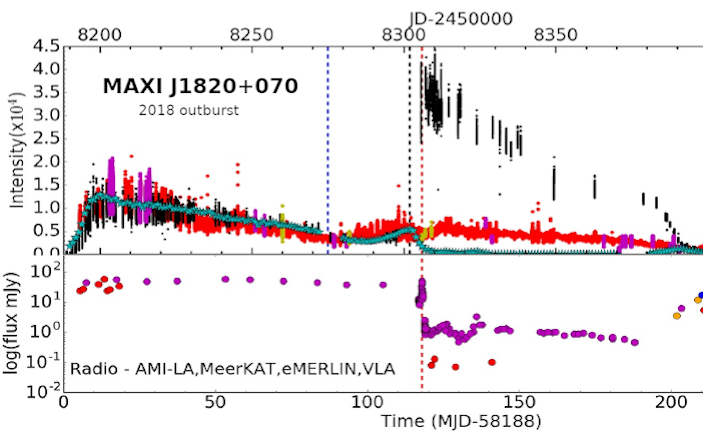}}
\caption{Overview of the soft (Swift/XRT - black) and hard (Swift/BAT - green) X-ray light-curves of J1820, along with the AAVSO optical (red).  The red dashed line marks the radio flare/jet ejection and X-ray state transition (from Hard to Soft). Adapted from \citep{Thomas22}.}
\label{fig:J1820overall-lc}
\end{subfigure}
\hfil
\begin{subfigure}{0.45\textwidth}
\scalebox{0.9}{\includegraphics{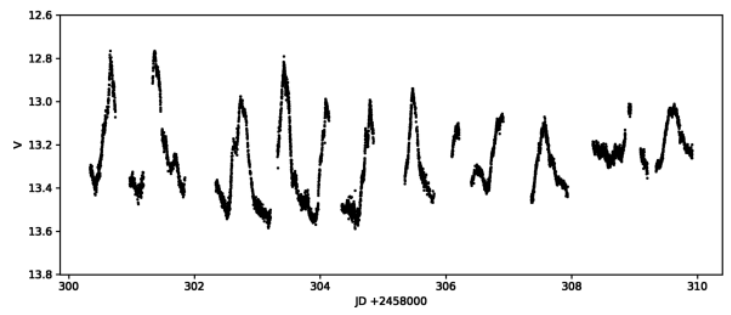}}
\caption{Zoom of the AAVSO optical light-curve of J1820 showing the large amplitude ``superhumps'' that start at day 87 (blue vertical line in (a)) \citep{Patterson19}.}
\label{fig:J1820SHs}
\end{subfigure}
\caption{X-ray and optical light-curves of MAXI\,J1820+070}
\end{figure}

By the time of MAXI\,J1820+070's discovery (hereafter J1820) in March 2018, there were multiple X-ray missions operating (Swift, MAXI, XMM, NuSTAR, NICER, AstroSAT, HXMT) that provided superb coverage of this extended ($>$6 months), luminous outburst (see \citep{Stiele20} for a detailed X-ray overview).  The Swift data are in Figure \ref{fig:J1820overall-lc}, where the Swift/BAT light-curve (in green) shows J1820 in the Hard state until day 118 (relative to MJD\,58188), when it undergoes a major radio flare/jet ejection (lower panel) and transitions to the Soft state.  In spite of X-ray dips being detected by XMM \citep{Kajava19}, indicating a likely high orbital inclination, none of these X-ray data revealed J1820's orbital period. That was to come from the extensive, and unprecedented, optical monitoring provided by the AAVSO, their overall light-curve being shown in red in Figure \ref{fig:J1820overall-lc} \citep{Thomas22}.

\begin{figure}[htb]
\centering
\setkeys{Gin}{width=\linewidth}
\begin{subfigure}{0.45\textwidth}
\scalebox{1.0}{\includegraphics{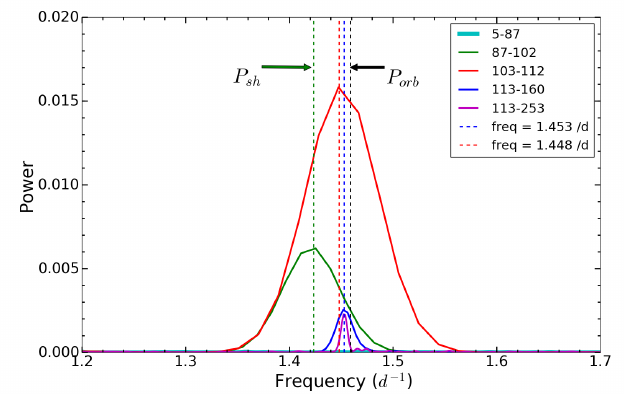}}
\caption{Lomb-Scargle periodograms of J1820 for different time intervals in the AAVSO light-curve (day numbers are MJD-58188) \citep{Thomas22}.}
\label{fig:J1820LS}
\end{subfigure}
\hfil
\begin{subfigure}{0.45\textwidth}
\scalebox{0.95}{\includegraphics{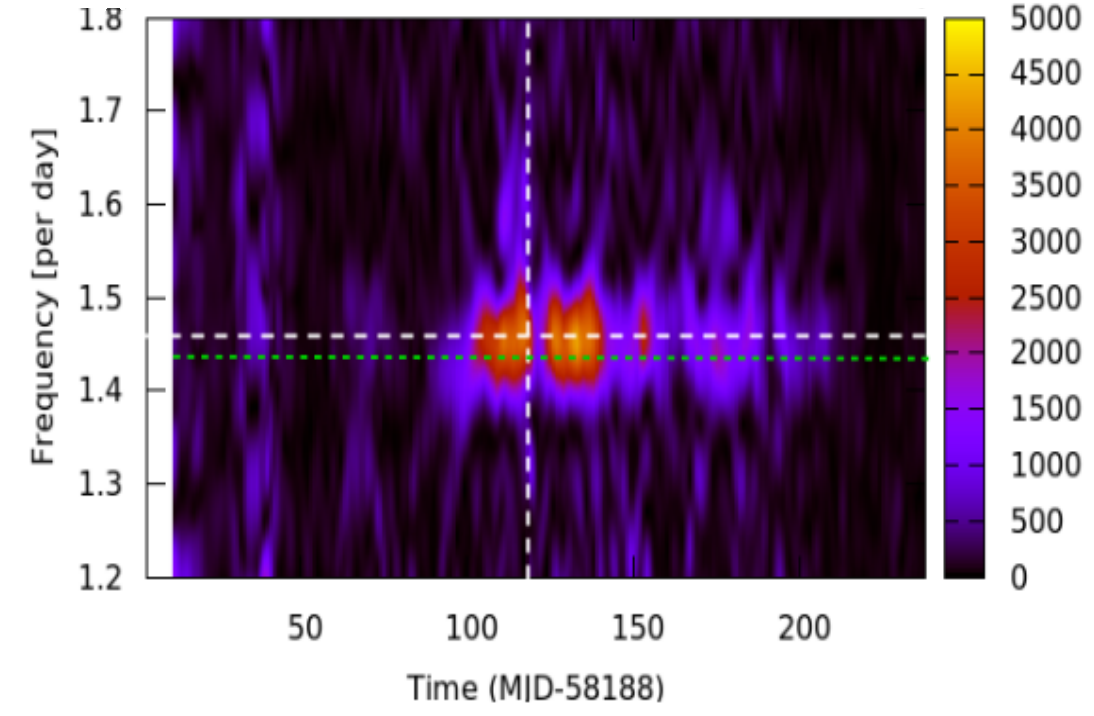}}
\caption{Dynamical LS power spectra (colour coded) of J1820 showing power turning on sharply around day 87.  White dashed lines mark the state transition (vertical) and  $P_{\rm orb}$ frequency (horizontal).  The green line marks the original $P_{\rm sh}$ value.  From \citep{Thomas22}.}
\label{fig:J1820DPS}
\end{subfigure}
\caption{Optical power spectra of MAXI\,J1820+070}
\end{figure}

\begin{figure}[htb]
\centering
\setkeys{Gin}{width=\linewidth}
\begin{subfigure}{0.4\textwidth}
\scalebox{0.8}{\includegraphics{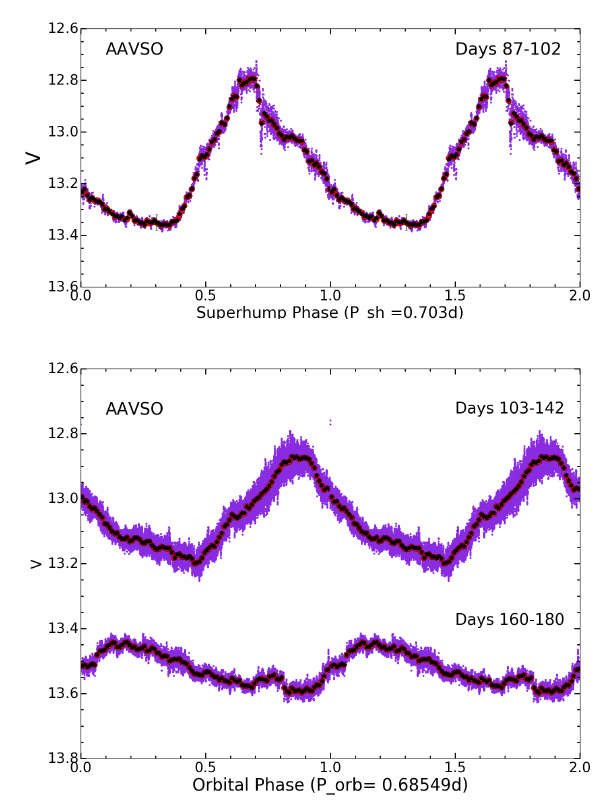}}
\caption{Folded optical light-curves of J1820 showing large ``superhump'' modulation (top box), and then orbital modulations (lower box) \citep{Thomas22}.}
\label{fig:J1820LCs}
\end{subfigure}
\hfil
\begin{subfigure}{0.5\textwidth}
\scalebox{1.0}{\includegraphics{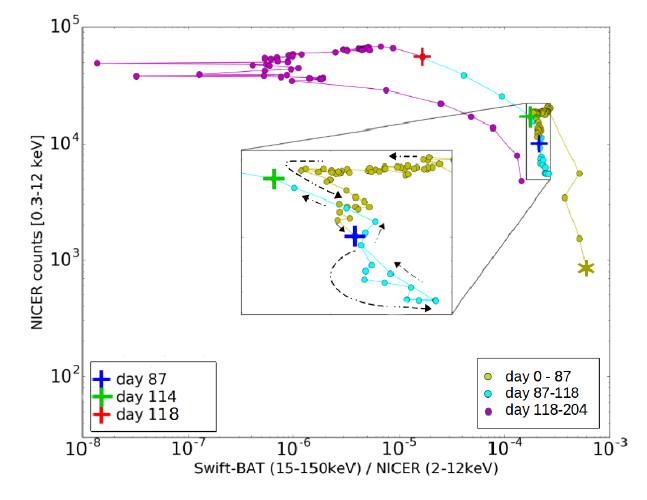}}
\caption{X-ray HID for J1820, with outburst starting at ``*'', and the interval approaching transition ({\color{red} {\bf +}}) is zoomed in the inset box, the large optical modulation begins at {\color{blue} {\bf +}}, and {\color{green}{\bf +}} is the Swift/BAT peak \citep{Thomas22}.}
\label{fig:J1820HID}
\end{subfigure}
\caption{Optical light curves and HID of MAXI\,J1820+070}
\end{figure}

However, the early months of outburst revealed no periods in the optical either, until day 87 when a large amplitude ($\geq$0.5\,mag) modulation suddenly became visible (Figure \ref{fig:J1820SHs}) at a period of 0.703\,d \citep{Patterson19}, with no apparent counterpart in X-rays. This continued for $\sim$2 weeks, but the period then gradually shortened as J1820 approached the state transition (see  Figure \ref{fig:J1820LS}), becoming close to, but never reaching, the $P_{\rm orb}$ of 0.6855\,d (determined subsequently spectroscopically in quiescence \citep{Torres20}).  This behaviour led \citep{Patterson19} to refer to the modulation as a ``superhump'', analogous to that seen in the SU\,UMa-subclass of CVs.  While these phenomena had been seen before in BHXTs \citep{DOD96}, their amplitudes were always much smaller (see Table \ref{tab:XRT-SHs-table}).  Furthermore, such long-term period changes had never been seen before, and are displayed via the ``dynamical power spectrum'' of Figure \ref{fig:J1820DPS}. Their large amplitude is demonstrated in the phase-folded light-curves (Figure \ref{fig:J1820LCs}), where the drift relative to orbital phase is also clearly visible.

{\small
\begin{table}
	\centering
	\caption{Black-hole X-ray transients displaying superhumps during outburst \citep{Thomas22}}
	\label{tab:XRT-SHs-table}
\begin{tabular}{l c c c c c}
\hline
{System name} & {$P_{\mathrm{sh}}$} (d)& {$P_{\mathrm{orb}}$} (d) & {$\epsilon$}(\%)$^1$  & {$A$} (mag)$^2$ & {$i$} (deg)$^3$\\
\hline
XTE~J1118+480 & 0.169930 & 0.170529 & 0.35 & 0.1 &71--82\\
GRO~J0422+32 & 0.2157 & 0.21216 & 1.67 & 0.1 & 10--50\\
GS~2000+25 & 0.3474 & 0.344098 & 0.96 & 0.2 & 54--60\\
GRS~1124-683 & 0.4376  & 0.4333 & 0.99 & 0.2 & 39--65\\
{\it MAXI~J1820+070} & {\it 0.70303} & {\it 0.68549} & {\it 2.6} & {\it 0.36} & {\it 63--81}\\
%\hline
%Swift~J1753.5-0127$^3$ & 0.1351 & ? & ? & 0.15 & ?\\
%GRS~1716-249$^4$ & 0.6127 & ? & ? & 0.1 & ?\\
\hline
{\footnotesize$^1$ $\epsilon = (P_{\rm sh}-P_{\rm orb}/P_{\rm orb}$}\\
{\footnotesize$^2$ $A$ = amplitude of photometric modulation}\\
{\footnotesize$^3$ $i$ = orbital inclination}
\end{tabular}
\end{table}
}

The timing of these events is also marked on the X-ray HID (hardness-intensity diagram, Figure \ref{fig:J1820HID}, where hardness is the Swift/BAT to NICER ratio), revealing that the optical modulation begins just as J1820 executes a spectral loop, first getting harder, then rapidly softer as the state transition approaches.  This is linked to the decreasing inner disc radius at this time (e.g. \citep{Kara19}, \citep{Zdziarski21}), as also revealed by the increasing X-ray QPO frequency \citep{Stiele20}, and accompanied by an increased height of the hard X-ray emitting region \citep{deMarco21}.  These are critical for explaining the huge amplitude of the optical modulation seen in J1820, which is unique amongst the BHXTs (Table \ref{tab:XRT-SHs-table}).  

\begin{figure}[htb]
\centering
\setkeys{Gin}{width=\linewidth}
\begin{subfigure}{0.45\textwidth}
\scalebox{1.0}{\includegraphics{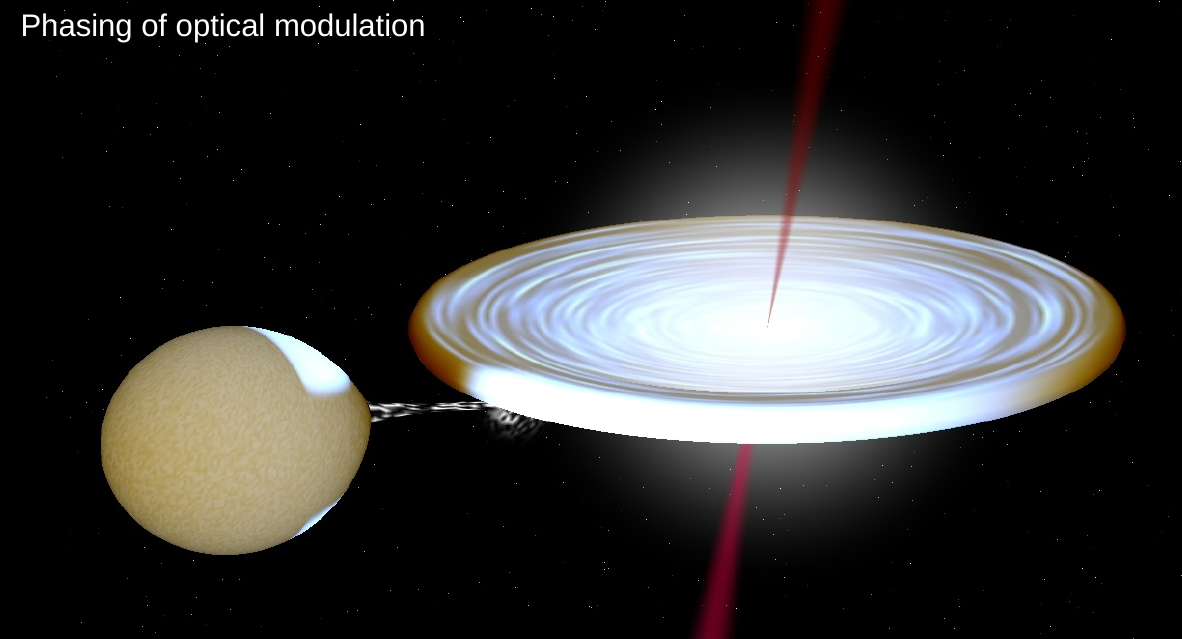}}
\caption{BinSim \footnote{\url{http://www.phys.lsu.edu/~rih/binsim/download.html}} visualisation of phasing of peak optical modulation in J1820 \citep{Thomas22}.}
\label{fig:J1820BinSim}
\end{subfigure}
\hfil
\begin{subfigure}{0.45\textwidth}
\scalebox{0.9}{\includegraphics{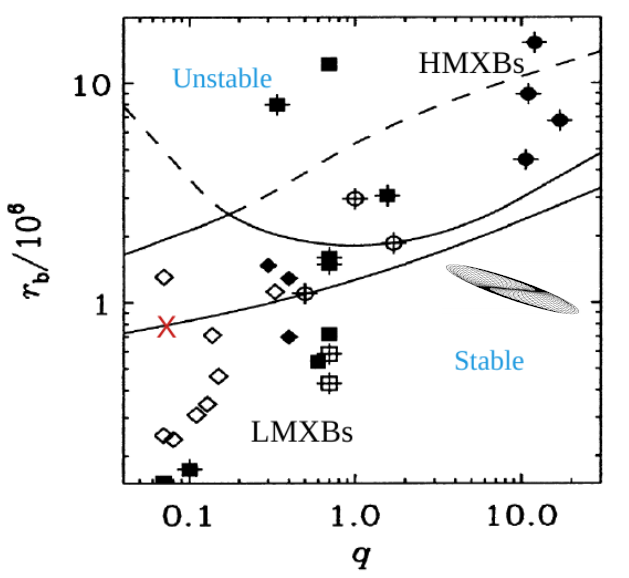}}
\caption{Stability diagram against radiation-driven disc warping for XRBs \citep{Ogilvie01}.  J1820 is marked by {\color{red} {\bf X}}. }
\label{fig:J1820-ODstability}
\end{subfigure}
\caption{Visualisation of MAXI\,J1820+070 and XRB stability diagram}
\end{figure}

The peak of this modulation occurs around orbital phase 0.85 (orientation shown in Figure \ref{fig:J1820BinSim}), and cannot therefore be due to the irradiated inner face of the donor, nor can it be due to the hot spot produced by the stream impact region as the phasing drifts and the mass transfer required would be unfeasibly high.  Nor can it be an X-ray modulation on this period as none is seen. And it certainly cannot be due to an eccentric disc \citep{Haswell01} as this can barely produce a 0.1\,mag modulation, let alone an amplitude more than 5$\times$ greater. 

Instead \citep{Thomas22} invoke the radiation-driven warping scenario \citep{Ogilvie01}, developed to account for accretion disc warping that was first seen in Her\,X-1 in the 1970s.  They showed that a system's stability is a function of the binary separation, $r_b$, and the mass ratio, $q$, as shown in Figure \ref{fig:J1820-ODstability}.  J1820 is right on the stability boundary, indicating that such a disc warp could occur, thereby increasing the projected surface area for an observer, and together with a raised hard X-ray emitting region, this could then produce the large optical modulation seen from day~87 onwards.  The slight period drift of the modulation also makes an irradiation-driven warp more likely as such a warp is likely to slowly precess.  However, such a model requires a binary inclination closer to the lower limit of the range in Table \ref{tab:XRT-SHs-table}, i.e. $\sim$63$^\circ$, so that the warp does not eclipse the X-ray source.  Interestingly, this is the inclination derived for the radio jet axis \citep{Atri20}.

\section*{Acknowledgments}
PAC would like to thank the HEASA organisers for the invitation to give this review, and the anonymous referee for a very careful reading of the original manuscript.
The SALT observations described here were obtained under the SALT Large Science Programme on transients (2018-2-LSP-001; PI: DAHB).  The SA authors acknowledge research support from the National Research Foundation. JPL was supported by a grant from the Space Agency CNES.

\bibliographystyle{unsrt}
%\bibliography{bibliography-HEASA}

\end{document}